\documentclass{ws-procs975x65}

\usepackage{wasysym}
\usepackage{units}

\begin{document}

\title{2-D MHD CONFIGURATIONS FOR ACCRETION DISKS AROUND MAGNETIZED STARS}

\author{R. BENINI}
\address{ICRA -- International Center for Relativistic Astrophysics\\
Dep. of Physics - ``Sapienza'' Universit\`a di Roma, Piazza A. Moro, 5 (00185), Roma, Italy\\
E-mail: riccardo.benini@icra.it}

%

%

\author{G. MONTANI}
\address{ENEA -- C.R. Frascati (Unit\`a Fus. Mag), Via E. Fermi, 45 (00044), Frascati (RM), Italy\\
ICRANet -- C. C. Pescara, Piazzale della Repubblica, 10 (65100), Pescara, Italy\\
E-mail: montani@icra.it}

We discuss basic features of steady accretion disk morphology around magnetized compact astrophysical objects. A comparison between the standard model of accretion based on visco-resistive MHD and the plasma instabilities, like ballooning modes, triggered by very low value of resistivity, is proposed.

\paragraph{Basic Model.}
The standard model of an axisymmetric thin disk
is fixed by the fluid-dynamical equilibria
established within the gravitational field
of the compact central object (having mass $M_S$)
when averaged out of the vertical direction \cite{B01}.
The angular frequency
of the disk takes the Keplerian profile
$\omega (r)=\omega _K=\sqrt{GM_S/r^3}$. Significant
deviations from such a behavior are expected only
in advective dominated regimes. The vertical equilibrium
gives, for an isothermal disk of temperature $T$,
the exponential decay of the mass density over the
equatorial plane value $\epsilon _0(r)$, i. e.
$D(z^2)\equiv \epsilon /\epsilon _0 =
\exp \{ -z^2/H^2\}$, with
$H = \sqrt{2v_s^2/\omega _K^2}$ being the half-depth
of the disk and $v_s$ the sound velocity
on the equatorial plane, that is to say
$v_s^2=2K_B T/m_i$ (here $K_B$ denotes the
Boltzmann constant, while $m_i$ stands for
the ions mass).
The azimuthal equilibrium describes the angular
momentum transport across the disk, by virtue of
a turbulent viscosity coefficient $D^0_{\mu}$, i. e.
${\dot{M}}_d(L - L_{0}) = 3\pi D^0_{\mu} \omega _Kr^2$,
where $L$ is the angular momentum per unit mass,
$L_{0}$ a fixed value and
$\dot{M}_d=-2\pi r\sigma v_r$ is the mass accretion
rate, associated to the radial velocity $v_r<0$ and to the
surface mass density $\sigma \equiv \int_{-H}^{H}\epsilon dz$.
Finally the continuity equation implies that
${\dot{M}}_d = \textrm{const}>0$.

\paragraph{The viscosity Coefficient.}

The viscosity coefficient $D^0_{\mu}$ that arises in the disk, as estimated by the microscopic plasma structure, 
results to be too small to account for the accretion rates observed
in some astrophysical systems, like X-ray binaries.
In fact, the observed accretion rates, 
evaluated by the increasing disk luminosity
${\dot{L}}_d\sim G {\dot{M}}_d M_S/R_{S}$,
require a large value of $D^0_{\mu}$, that in \cite{S73}
were postulated to be due to a turbulent behavior
of the disk plasma. Since, by definition
$L=\omega r^2$, we can infer that 
$D^0_{\mu} = 2\sigma v_tH/3$, $v_t$ being a turbulence
velocity expressible as $v_t=\alpha v_s$,
$\alpha$ being a free parameter. The point is
that the axisymmetric disk is linearly stable with
respect to small perturbations preserving its symmetry.
For a discussion of the onset of turbulence by
magnetohydrodynamics (MHD) instabilities
based on the Velikov analysis of 1959, see
the review article \cite{B98}.
In what follows, we address a different point of view, based on the idea that the accretion phenomenon is related to the structures rising within the plasma disk.

\paragraph{MHD Disk Configuration.}

The magnetic field, characterizing the central
object, takes the form 
$r \vec{B} = \partial _z\psi \vec{e}_r +
{I}\vec{e}_{\phi } + 
\partial _r\psi \vec{e}_z$
with $\psi = \psi (r\, ,z^2)$ and
$I = I(\psi \, , z)$.
The matter flux within the disk reads as
\begin{equation}
\epsilon \vec{v} =
-\frac{1}{r}\partial _z\Theta \vec{e}_r +
\epsilon \omega (r\, ,z^2)r\vec{e}_{\phi } +
\frac{1}{r}\partial _r\Theta \vec{e}_z
\, ,
\label{solconteq}
\end{equation}
where $\Theta (r\, , z)$ has to be an odd function of $z$
to deal with a non-zero accretion rate.

A local model of the equilibrium \cite{C05,CR06,BM09},
around a radius value $r = r_0$,
 allows to study
the effects of the electromagnetic reaction
of the disk plasma.
We split the energy density and
the pressure as
$\epsilon = \bar{\epsilon }(r_0,\, z^2) + \hat{\epsilon}$
and 
$p = \bar{p}(r_0,\, z^2) + \hat{p}$, respectively. 
As well, the magnetic surface function
splits in the form 
$\psi = \psi _0(r_0) +
\psi _1(r_0\, , r-r_0\, ,z^2)$, with $\psi _1\ll \psi _0$.
The corotation theorem \cite{F37} states  that
$\omega = \omega (\psi )$.
Hence, we can take the decomposition 
$\omega = \omega _K +
\omega ^{\prime }_0\psi _1$,
where $\omega ^{\prime }_0 = \textrm{const}$.
We now define the dimensionless functions 
$Y$, $\hat{D}$ and $\hat{P}$, as follows
\begin{equation}
Y\equiv \frac{k_0\psi _1}{\partial _{r_0}\psi _0}
\, , \,\hspace{5mm}
\hat{D}\equiv \frac{\beta \hat{\epsilon}}{\epsilon _0}
\, , \,\hspace{5mm}
\hat{P}\equiv \beta \frac{\hat{p}}{p_0}
\, ,
\label{deff}
\end{equation}
where $p_0\equiv 2K_B\hat{T}\epsilon _0/m_i$
and $\beta \equiv 8\pi p_0/B^2_{0z} = 
1/(3\epsilon _z^2) \equiv k_0^2H_0^2/3$.
Here  $k_0$ is the fundamental wavenumber
of the radial equilibrium, defined as
$k_0\equiv 3\omega _K^2/v_A^2$, with
$v_A^2\equiv 4\pi \epsilon _0/B^2_{z0}$, recalling
that $B_{z0} = \partial _{r_0}\psi _0/r_0$.
Thus, we deal with the dimensionless
radial and vertical variables $x\equiv k_0(r - r_0)$ and  $u\equiv z/(\sqrt{\epsilon _z}H_0)$, respectively.
The vertical and radial equilibria can be restated as follows
\begin{eqnarray}
&\partial _{u^2}\hat{P} + \epsilon _z\hat{D}
+ 2\left(\partial ^2_{x^2}Y +
\epsilon _z\partial ^2_{u^2}Y\right)
\partial _{u^2}Y = 0\label{vertad}
\, ,&\\
&\left(D + \frac{1}{\beta }\hat{D}\right) Y + 
\partial ^2_{x^2}Y +
\epsilon _z\partial ^2_{u^2}Y        
+\frac{1}{2}\partial _x\hat{P} +
\left(\partial ^2_{x^2}Y +
\epsilon _z\partial ^2_{u^2}Y\right)
\partial_xY = 0 &
\, .
\label{radad}
\end{eqnarray}

\paragraph{The electron force balance equation.}

\begin{figure}[htbp] 
   \centering
   \includegraphics[width=.4\textwidth]{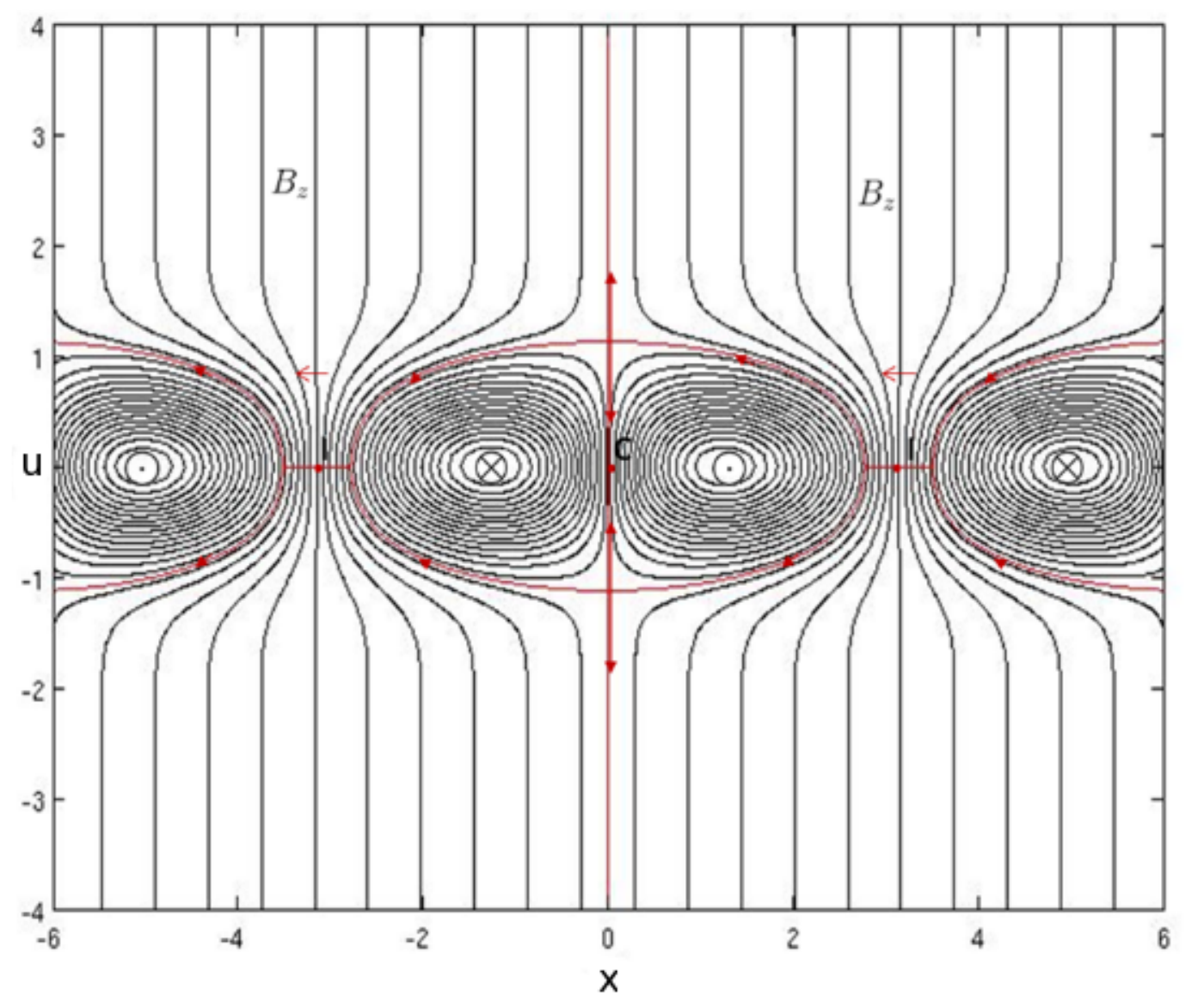} 
   \caption{This picture sketches the plasma along the separatrix.}
   \label{fig:example}
\end{figure}

In the presence of a non-zero resistivity coefficient
$\eta$, the equation of the electron force
balance, reads as:
$\vec{E} + \frac{\vec{v}}{c}\wedge \vec{B} =
\eta \vec{J}$, 
$\vec{E}$ denoting the electric field
and $\vec{J}$ the current density.
Since the axial symmetry requires
$E_{\phi }\equiv 0$,
in the local formulation around $r_0$,
the azimuthal component  stands as
\begin{equation}
v_rB_z - v_zB_r = 
\frac{\eta c^2}{4\pi r_0}
\left(\partial ^2_r\psi _1 + \partial_z^2\psi _1\right)
\, .
\label{efbloc}
\end{equation}
The discussion about the microscopic viscosity
coefficient can be directly extrapolated for the
resistivity coefficient too. Thus, for instance, in
the linear regime $B_r\sim 0$, the radial matter infall
associated to the dissipative term balancing the electron
equilibrium, can not provide a sufficient accretion rate.
Therefore, we are lead to argue that the azimuthal
electron force balance  stands on average
as $\langle v_rB_z - v_zB_r\rangle \simeq 0$.
The driving force of the mass accretion must be searched
in the peculiar MHD instabilities arising in the plasma
as triggered by very small resistivity contributions,
like the so-called \emph{resistive ballooning modes}\cite{C77}. 
Eqs. \ref{vertad}-\ref{radad} admit, at lowest order in $\epsilon_{z}$, a solution which gives the relevant magnetic surfaces in the form
\begin{equation}
Y_{t}\simeq x - Y^{0}_{0} \sin x  F_{0}(u^{2}),\, \hspace{5mm} Y^{0}_{0} = \textrm{const}
\end{equation}
We argue that in the region around $r_{0} + \pi/k_{0}$ the magnetic field lines are crowded between two adjacent separatrices and the flow across them can be thought as being maintained intermittently by recurrent resistive ballooning modes. Such modes,  driven by the combined effect of the local radial density gradient and of gravity, let the plasma slip through the magnetic field lines (as sketched in Fig. \ref{fig:example}).

\vspace{-3mm}
\vspace{-3mm}

This  work  was developed within the framework of the CGW Collaboration
({\sf www.cgwcollaboration.it)} and the authors would like to thank Bruno Coppi for his valuable advice on the topic and for providing us with the Fig. \ref{fig:example}.
%

\begin{thebibliography}{1}

\bibitem{B01}
G.~S. {Bisnovatyi-Kogan} and R.~V.~E. {Lovelace}, {\em New Astronomy Review}
  {\bf 45}, 663 (2001).

\bibitem{S73}
N.~I. {Shakura}, {\em Soviet Astronomy} {\bf 16}, 756 (1973).

\bibitem{B98}
S.~A. {Balbus} and J.~F. {Hawley}, {\em Reviews of Modern Physics} {\bf 70}, 1
  (1998).

\bibitem{C05}
B.~{Coppi}, {\em Physics of Plasmas} {\bf 12}, 057302 (2005).

\bibitem{CR06}
B.~{Coppi} and F.~{Rousseau}, {\em The Astrophysical Journal} {\bf 641}, 458
  (2006).

\bibitem{BM09}
G.~Montani and R.~Benini, {\em to appear on Modern Physics Letters A}  (2009).

\bibitem{F37}
V.~C.~A. {Ferraro}, {\em Monthly Notices of the Royal Astronomical Society}
  {\bf 97}, 458 (1937).


\bibitem{C77}
B.~Coppi, {\em Physical Review Letters} {\bf 39}, 939 (1977).

\end{thebibliography}

\end{document}